\documentclass[fleqn,twoside]{article}
\usepackage{espcrc2}

\usepackage{graphicx}
\usepackage[figuresright]{rotating}

\newcommand{\AmS}{{\protect\the\textfont2
  A\kern-.1667em\lower.5ex\hbox{M}\kern-.125emS}}
  
\newcommand{\lsim}{
\mathrel{\hbox{\rlap{\hbox{\lower4pt\hbox{$\sim$}}}\hbox{$<$}}}}

\title{In Pursuit of New Physics in the B System}

\author{R. Fleischer\address[MCSD]{Theory Division, Department of Physics,
CERN, CH-1211 Geneva 23, Switzerland}}
       
\begin{document}


\thispagestyle{empty}

\begin{flushright}
CERN-PH-TH/2006-139\\
hep-ph/0607241
\end{flushright}

\vspace{2.0truecm}
\begin{center}
\boldmath
\large\bf In Pursuit of New Physics in the B System
\unboldmath
\end{center}

\vspace{0.9truecm}
\begin{center}
Robert Fleischer\\[0.1cm]
{\sl Theory Division, Department of Physics, CERN\\
CH-1211 Geneva 23, Switzerland}
\end{center}

\vspace{0.9truecm}

\begin{center}
{\bf Abstract}
\end{center}

{\small
\vspace{0.2cm}\noindent
The $B$-meson system offers interesting probes for the search
of physics beyond the Standard Model. After addressing possible
signals of new-physics contributions to the $B\to\phi K$ and
$B\to\pi K$ decay amplitudes, we focus on the data for
$B^0_q$--$\bar B^0_q$ mixing ($q\in\{d,s\}$), giving a critical
discussion of their interpretation in terms of model-independent
new-physics parameters. We address, in particular, the impact of
the uncertainties of the relevant input parameters, discuss benchmarks
for future precision measurements at the LHC, and explore the prospects 
for new CP-violating effects in the $B_s$-meson system, which could be 
detected at the LHC.
}

\vspace{0.9truecm}

\begin{center}
{\sl Invited talk at the 1st Workshop on Theory, Phenomenology and Experiments 
in Heavy Flavour Physics\\
Anacapri, Capri, Italy, 29--31 May 2006\\
To appear in the Proceedings (Nucl.\ Phys.\ B: Proc.\ Suppl., Elsevier)}
\end{center}

\vfill
\noindent
CERN-PH-TH/2006-139\\
July 2006

\newpage
\thispagestyle{empty}
\vbox{}
\newpage
 
\setcounter{page}{1}


\begin{abstract}
The $B$-meson system offers interesting probes for the search
of physics beyond the Standard Model. After addressing possible
signals of new-physics contributions to the $B\to\phi K$ and
$B\to\pi K$ decay amplitudes, we focus on the data for
$B^0_q$--$\bar B^0_q$ mixing ($q\in\{d,s\}$), giving a critical
discussion of their interpretation in terms of model-independent
new-physics parameters. We address, in particular, the impact of
the uncertainties of the relevant input parameters, discuss benchmarks
for future precision measurements at the LHC, and explore the prospects 
for new CP-violating effects in the $B_s$-meson system, which could be 
detected at the LHC.
\vspace{1pc}
\end{abstract}

\maketitle

\section{SETTING THE STAGE}
Thanks to the $B$ factories, remarkable progress in
the testing of the Kobayashi--Maskawa  \cite{KM} mechanism of CP violation
could be made over the recent years. The analyses of unitarity triangle of the
Cabibbo--Kobayashi--Maskawa (CKM) matrix by the CKMfitter \cite{CKMfitter} 
and UTfit \cite{UTfit} collaborations show impressive global agreement with the 
picture of the Standard Model (SM), although it is no longer perfect, with some 
tension in the corresponding plots and correlations. 

The theoretical tool for the description of weak decays and particle--antiparticle
mixing is given by low-energy effective Hamiltonians. In this framework,
the heavy degrees of freedom are ``integrated out", and are encoded in 
perturbatively calculable Wilson coefficients $C_k(\mu)$. 
On the other hand, the long-distance physics resides in the hadronic matrix 
elements of local operators. 

In this formalism, there are two possibilities for new physics (NP) to enter:
(i) modification of $C_k(\mu=M_W)\to C_k^{\rm SM}+C_k^{\rm NP}$,
where the NP pieces $C_k^{\rm NP}$ may involve new 
CP-violating phases; (ii) new operators may enter the stage:
$\{Q_k\}\to\{Q_k^{\rm SM}, Q_l^{\rm NP}\}$,
involving, in general, new sources of flavour and CP violation.

Many specific NP analyses can be found in the literature.
In general, they suffer from the problem that the choice of the NP model is 
governed by personal ``biases", and that the predictivity is inversely proportional 
to the number of NP parameters. However, the central problem for the 
resolution of  NP effects in weak processes is related to hadronic uncertainties. 
Concerning non-leptonic $B$ decays, interesting progress could recently be 
made, as discussed in the talks by A. Khodjamirian and Z. Ligeti, although the 
data show that the theoretical challenge remains. 

Fortunately, the $B$-meson system offers various powerful strategies to circumvent 
the calculation of the hadronic matrix elements that can be implemented at
the $B$ factories and later on at the LHC. In these methods, amplitude relations 
(exact or derived from QCD flavour symmetries) or interference effects in decays 
of neutral $B_q$ mesons ($q\in\{d,s\}$), which may lead to 
mixing-induced CP violation, are used \cite{RF-JPHYSG}. Following these lines, 
a rich roadmap for the exploration of quark-flavour physics emerges, where popular avenues for NP to enter are given by effects at the decay amplitude level or in
$B^0_q$--$\bar B^0_q$ mixing.

\section{NEW PHYSICS IN AMPLITUDES}
At the decay-amplitude level, NP effects are small if tree processes of the SM play 
the dominant role. On the other hand, there are potentially large effects in
flavour-changing neutral-current processes through new particles in loop diagrams
or  new contributions at the tree level. In the following discussion, we 
address two possible signals for this kind of NP contributions: the $B\to\phi K$ and 
$B\to\pi K$ systems.

\boldmath
\subsection{Challenging the SM through $B\to\phi K$}
\unboldmath
The $B^0_d\to\phi K_{\rm S}$ mode is a $\bar b\to\bar s$ penguin process,
which is dominated by QCD penguins, but receives also significant  
electroweak (EW) penguin contributions. If we neglect corrections of 
${\cal O}(\lambda^2)$ in the Wolfenstein parameter $\lambda\equiv|V_{us}|=0.22$, 
the CKM structure of the $B^0_d\to\phi K_{\rm S}$ channel implies the following 
relations for the direct and mixing-induced CP asymmetries:
\begin{eqnarray}
{\cal A}_{\rm CP}^{\rm dir}(B_d\to\phi K_{\rm S})&=&0
\\
\underbrace{{\cal A}_{\rm CP}^{\rm mix}(B_d\to\phi K_{\rm S})}_{\equiv
-(\sin2\beta)_{\phi K_{\rm S}}}&=&
\underbrace{{\cal A}_{\rm CP}^{\rm mix}(B_d\to\psi K_{\rm S})}_{\equiv
-(\sin2\beta)_{\psi K_{\rm S}}}.
\end{eqnarray}
The penguin decay $B^0_d\to\phi K_{\rm S}$ is a sensitive probe for NP, 
so that these relations may well be violated through its impact. 
During the last years, the BaBar and Belle data have converged,
yielding now the following averages \cite{HFAG}:
\begin{eqnarray}
{\cal A}_{\rm CP}^{\rm dir}(B_d\to \phi K_{\rm S})&=&-0.09\pm0.14\\
(\sin2\beta)_{\phi K_{\rm S}}&=&0.47\pm0.19,\label{s2b-phiK}
\end{eqnarray}
so that ${\cal S}_{\phi K}\equiv (\sin 2\beta)_{\phi K_{\rm S}}- 
(\sin 2\beta)_{\psi K_{\rm S}}=-0.22\pm0.19$. The central value
of this result could generically be accommodated in extensions of
the SM, although the current experimental errors are too large to
draw definite conclusions. In order to get the whole picture, also
$B^\pm\to\phi K^\pm$ observables are required. Such an analysis
shows that the $B\to\phi K$ data may indicate a modified EW penguin sector
with a large CP-violating phase \cite{RF-JPHYSG}, complementing
the  ``$B\to\pi K$ puzzle".

\boldmath
\subsection{Challenging the SM through $B\to\pi K$}
\begin{figure}
\begin{center}
\includegraphics[width=6.4truecm]{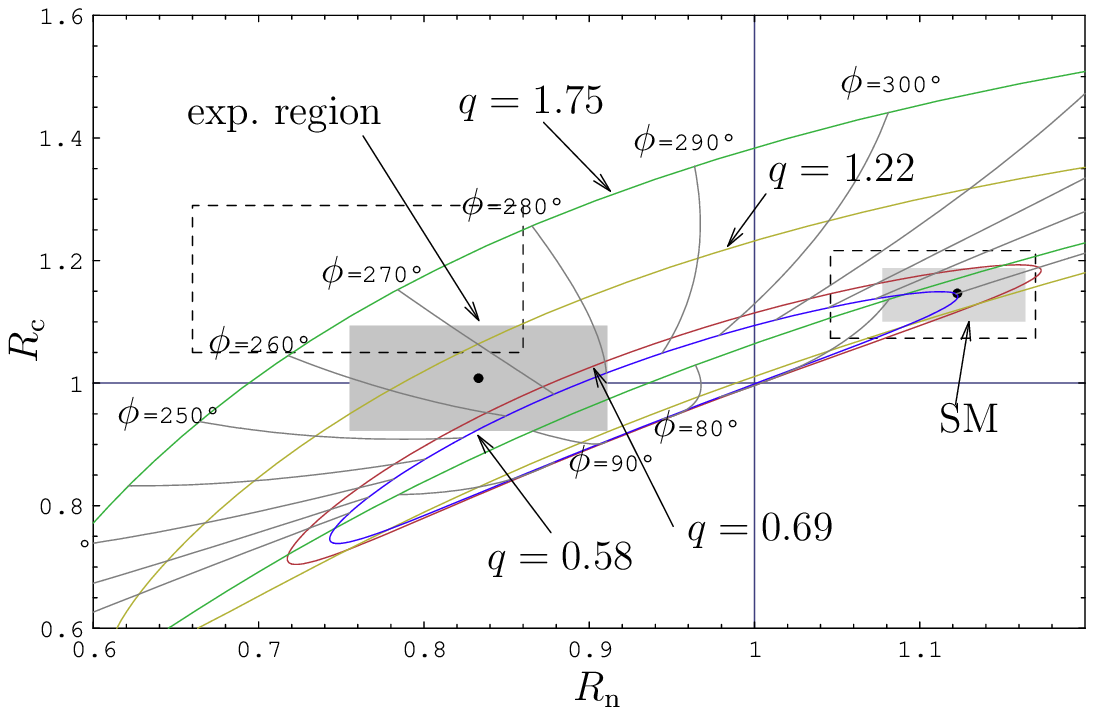}
\end{center}
\vspace*{-1.2truecm}
\caption{The situation in the $R_{\rm n}$--$R_{\rm c}$ plane.}\label{fig:RnRc}
\end{figure}

\unboldmath
There is a long history of studies of the these decays, and since
the year 2000, the $B$-factory data raise the question of a
discrepancy with the SM. This ``$B\to\pi K$ puzzle" is reflected by 
the observables of those decays that are significantly affected by EW penguins,
and  the following ratios are of central interest:
\begin{eqnarray}
R_{\rm c}&\equiv&2\left[\frac{\mbox{BR}(B^\pm\to\pi^0K^\pm)}{\mbox{BR}
(B^\pm\to\pi^\pm K)}\right]
\\
R_{\rm n}&\equiv&\frac{1}{2}\left[\frac{\mbox{BR}(B_d\to\pi^\mp K^\pm)}{\mbox{BR}
(B_d\to\pi^0K)}\right].
\end{eqnarray}
Here the  EW penguin effects enter through the decays with neutral 
pions, and are described  both by a parameter $q$, which measures the strength 
of the EW penguins with respect to the tree topologies, and by a CP-violating phase 
$\phi$. In the SM, this phase vanishes, and $q$ can be calculated with the help of the 
$SU(3)$ flavour symmetry. In Ref.~\cite{BFRS}, where also a comprehensive
guide to the $B\to\pi K$ literature can be found, a systematic strategy for the
exploration of the $B\to\pi K$ system was developed. It uses the $B$-factory
data for $B\to\pi\pi$ decays as the starting point, and determines then
the hadronic $B\to\pi K$ parameters with the help of the $SU(3)$ flavour
symmetry. The resulting situation can transparently be discussed in the 
$R_{\rm n}$--$R_{\rm c}$ plane, as shown in Fig.~\ref{fig:RnRc} for the
current data \cite{BFRS-05}: the shaded areas indicate the SM prediction and
the experimental range, the lines show the theory predictions for the central values 
of the hadronic parameters and various values of $q$ with $\phi\in[0^\circ,360^\circ]$;
the dashed rectangles represent the SM predictions and experimental ranges 
at the time of the original analysis of Ref.~\cite{BFRS}.  Although the central values 
of $R_{\rm n}$ and $R_{\rm c}$ have slightly moved towards each other, the 
puzzle is as prominent as ever. The experimental region can now be reached 
without an enhancement of $q$, but a large CP-violating phase $\phi$ of the order 
of $-90^\circ$ is still required, although $\phi\sim+90^\circ$ can also bring us rather 
close to the experimental range of $R_{\rm n}$ and $R_{\rm c}$.

Following Ref.~\cite{BFRS}, also the CP asymmetries of the 
$B^\pm\to\pi^0 K^\pm$ and $B_d\to\pi^0K_{\rm S}$ modes can be predicted
both in the SM and in the scenario of NP effects in the EW penguin sector. 
The mixing-induced CP asymmetry of the latter decay has recently received a 
lot of attention, as the current $B$-factory data give an experimental 
value of $-0.38\pm 0.26$ for the quantity
\begin{equation}
\Delta S \equiv (\sin2\beta)_{\pi^0K_{\rm S}}-(\sin2\beta)_{\psi K_{\rm S}}.
\end{equation}
In the strategy described above, this difference is predicted to be {\it positive}
in the SM, and in the ballpark of $0.10$--$0.15$. Interestingly,
the best values for $(q,\phi)$ that are implied by the measurements of $R_{\rm n,c}$ 
make the disagreement of $\Delta S$ with the data even larger than in the SM. 
However, also values of $(q,\phi)$ can be found for which $\Delta S$ could be 
smaller than in the SM or even reverse the sign. This happens in
particular for $\phi\sim+90^\circ$, i.e.\ if the CP-violating NP phase flips its sign. 
In this case, also the central value of $(\sin2\beta)_{\phi K_{\rm S}}$ in (\ref{s2b-phiK})
could be straightforwardly accommodated in this scenario of NP \cite{RF-JPHYSG},
and could in fact be another manifestation of a modified EW penguin sector
with new sources for CP violation.

\begin{figure}
\begin{center}
\includegraphics[width=6.4truecm]{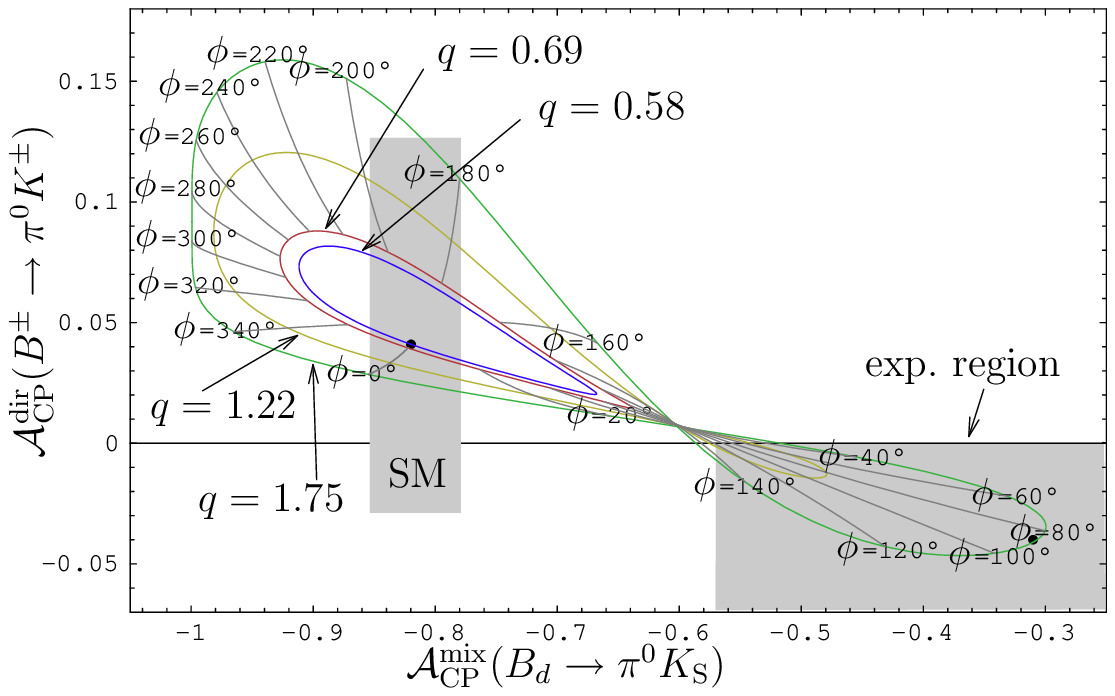}
\end{center}
\vspace*{-1.2truecm}
\caption{Illustration of the situation in the 
${\cal A}_{\rm CP}^{\rm mix}(B_d\to\pi^0K_{\rm S})$--${\cal A}_{\rm CP}^{\rm mix}
(B^\pm\to\pi^0K^\pm)$ plane.}\label{fig:ACP}
\end{figure}

Another fruitful testing ground for this scenario is given by the interplay with
rare decays such as $K^+\to\pi^+\nu\bar\nu$, $K_{\rm L}\to \pi^0\nu\bar\nu$ 
and $B_{s,d}\to\mu^+\mu^-$, which are sensitive probes of $Z$ penguins. 
As discussed in Ref.~\cite{BFRS-05}, the corresponding observables show 
specific patterns for NP scenarios satisfying the $B$-factory constraints from 
$B\to X_s\ell^+\ell^-$ decays, thereby allowing us to pin down a modified 
EW penguin sector. In these explorations, $K_{\rm L}\to \pi^0\nu\bar\nu$
turns out to be particularly interesting, as its branching ratio may be dramatically
enhanced. Further details of this strategy are discussed in the talk by F. Schwab.

\section{NEW PHYSICS IN B MIXING}
While $B^0_d$--$\bar B^0_d$ mixing is well established and
$\Delta M_d = (0.507\pm 0.004)\,{\rm ps}^{-1}$ known with impressive
experimental accuracy,  only lower bounds on $\Delta M_s$ 
were available,  for many years, from the LEP (CERN) 
experiments and SLD (SLAC). This spring, $\Delta M_s$ could eventually 
be pinned down at the Tevatron: D0  reported a two-sided bound 
$17 \,{\rm ps}^{-1}< \Delta M_s < 21\,{\rm ps}^{-1}$ (90\% C.L.),
corresponding to a 2.5\,$\sigma$ signal at $\Delta M_s=19\,{\rm ps}^{-1}$ 
\cite{D0}, and CDF announced the following result \cite{CDF}:
\begin{equation}
\Delta M_s = \left[17.31^{+0.33}_{-0.18}({\rm
  stat})\pm 0.07({\rm syst})\right]{\rm ps}^{-1}.
\end{equation}
These new measurements have already triggered considerable theoretical activity;
in the following discussion, we shall focus on the analysis of Ref.~\cite{BF-DMs},
where also a comprehensive guide to the recent literature on $\Delta M_s$
can be found.

\boldmath
\subsection{A Closer Look at $B^0_q$--$\bar B^0_q$ Mixing}
\unboldmath
If we use an effective Hamiltonian to write
\begin{equation}
\langle B_q^0| {\cal H}^{\Delta B=2}_{\rm eff} | \bar B_q^0\rangle = 2 M_{B_q}
M_{12}^q,
\end{equation}
the mass difference $\Delta M_q$ and the CP-violating mixing phase $\phi_q$
take the forms
\begin{equation}
\Delta M_q = 2 |M_{12}^q| \quad \mbox{and} \quad \phi_q = \arg (M_{12}^q).
\end{equation}
In the SM, $B^0_q$--$\bar B^0_q$ mixing arises
from box diagrams with top-quark exchanges, yielding
\begin{eqnarray}
\lefteqn{M_{12}^{q,{\rm SM}} = 
\frac{G_{\rm F}^2M_W^2}{12\pi^2} M_{B_q} (V_{tq}^\ast V_{tb})^2}\nonumber\\ 
&& \times S_0(x_t)\hat{\eta}^{B}\hat B_{B_q}f_{B_q}^2.\label{M12-SM}
\end{eqnarray}
The phases of the CKM factors imply
\begin{equation}
\phi_d^{\rm SM}=2\beta, \quad \phi_s^{\rm SM}=
-2\lambda^2\eta.
\end{equation}
In order to determine $|V_{tq}^\ast V_{tb}|$, which are required for the SM
predictions of $\Delta M_q$, we use the unitarity of the CKM matrix to express
them in terms of $|V_{cb}|$, the side 
$R_b\propto |V_{ub}/V_{cb}|$ of the unitarity triangle, and its angle $\gamma$. 
These quantities can be determined from tree-level processes, which are very 
robust with respect to NP effects. While $|V_{cb}| = (42.0\pm 0.7)\times 10^{-3}$, 
the situation of 
\begin{eqnarray}
|V_{ub}|_{\rm incl} &=& (4.4\pm 0.3)
\times 10^{-3}\label{Vub-incl}\\
|V_{ub}|_{\rm excl} &=& 
(3.8\pm 0.6)\times 10^{-3}\label{Vub-excl}
\end{eqnarray}
has to be clarified since inclusive and exclusive $b\to u\ell\bar \nu_\ell$ processes
show a discrepancy at the $1\sigma$ level. For a benchmark scenario of the year
2010, we assume that the inclusive value will be confirmed, with an error of 
$\pm0.2\times 10^{-3}$. Concerning $\gamma$, we use a value of $(65\pm 20)^\circ$,
and assume that it will move to $(70\pm5)^\circ$ in 2010 thanks to LHCb.

In (\ref{M12-SM}), the short-distance physics is described by the 
Inami--Lim function $S_0(x_t\equiv m_t^2/M_W^2)$
and the perturbative QCD factor $\hat{\eta}^{B}$, which are known quantities.
On the other hand, the long-distance physics is encoded in
$\hat B_{B_q}f_{B_q}^2$, which can be determined through lattice QCD studies.
Here the front runners are unquenched calculations with 2 or 3
dynamical quarks and Wilson or staggered light quarks, respectively. Despite 
tremendous progress, the results still suffer from several uncertainties. For the
following analysis, we use two sets of parameters from the JLQCD and
HPQCD collaborations:
\begin{eqnarray}
\left.\frac{f_{B_d}\hat{B}_{B_d}^{1/2}}{\rm GeV}\right|_{\rm JLQCD} &=& 0.215\pm
0.019^{+0}_{-0.023}\\
\left.\frac{f_{B_s}\hat{B}_{B_s}^{1/2}}{\rm GeV}\right|_{\rm JLQCD} &=& 0.245\pm
0.021^{+0.003}_{-0.002}\\
\xi_{\rm JLQCD}& = & 1.14\pm 0.06^{+0.13}_{-0},
\end{eqnarray}
which were obtained for two flavours of dynamical light Wilson quarks, and
\begin{eqnarray}
\left.\frac{f_{B_d}\hat{B}_{B_d}^{1/2}}{\rm GeV}\right|_{\rm (HP+JL)QCD}& =& 
0.244\pm 0.026\label{HPJL-d}\\
\left.\frac{f_{B_s}\hat{B}_{B_s}^{1/2}}{\rm GeV}\right|_{\rm (HP+JL)QCD} &=&
0.295\pm0.036\\
\xi_{\rm (HP+JL)QCD} & = &
1.210^{+0.047}_{-0.035},\label{HPJL-xi}
\end{eqnarray}
where $f_{B_q}$ comes from HPQCD (3 dynamical
flavours) and $\hat B_{B_q}$ from JLQCD as no value for this parameter
is available from the former collaboration; as usual, 
$\xi \equiv f_{B_s}\hat{B}_{B_s}^{1/2}/(f_{B_d}\hat{B}_{B_d}^{1/2})$.
For our 2010 scenario, we assume the set of the numerical values in 
(\ref{HPJL-d})--(\ref{HPJL-xi}).

\boldmath
\subsection{Space for NP in the $B_d$ System}
\unboldmath
In the presence of NP, $M_{12}^d$ can be written in the following model-independent
way:
\begin{equation}
M_{12}^d=M_{12}^{d,{\rm SM}}\left[1 + \kappa_d e^{i\sigma_d}\right].
\end{equation}
If we introduce a parameter $\rho_d$ through
\begin{equation}\label{rhod-def}
\rho_d\equiv
\left|\frac{\Delta M_d}{\Delta M_d^{\rm SM}}\right|=
\sqrt{1+2\kappa_d\cos\sigma_d+\kappa_d^2},
\end{equation}
the experimental result for $\Delta M_d$ and the theoretical prediction 
$ \Delta M_d^{\rm SM}$ allow us to determine $\kappa_d$ as a function of
the CP-violating phase $\sigma_d$, i.e.\ to constrain the space of the NP
parameters. Using the input parameters as specified above with
$\Delta M_d = (0.507\pm 0.004)\,{\rm ps}^{-1}$, we obtain
\begin{eqnarray}
\left.\rho_d\right|_{\rm JLQCD} &=&
  0.97\pm0.33^{-0.17}_{+0.26}\label{rhod-JLQCD}\\
\left.\rho_d\right|_{\rm  (HP+JL)QCD} &=& 
0.75\pm0.25\pm0.16,\label{rhod-HPJL}
\end{eqnarray}
where the first and second errors are due to $\gamma$ (and a small extent to
$R_b$) and $f_{B_d}
 \hat B_{B_d}^{1/2}$, respectively.

\begin{figure}
\begin{center}
\includegraphics[width=4.8truecm]{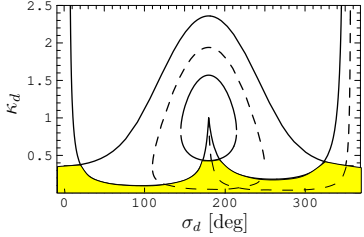}
\end{center}
\vspace*{-1.2truecm}
\caption{Currently allowed region in the $\sigma_d$--$\kappa_d$ plane 
for the JLQCD results and $\phi^{\rm NP}_d|_{\rm excl}$.}\label{fig:Bd-JLQCD}
\end{figure}

\begin{figure}
\begin{center}
\includegraphics[width=4.8truecm]{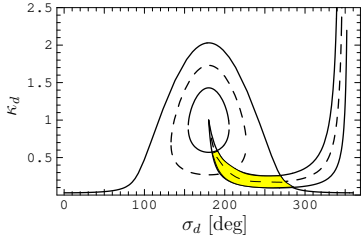}
\end{center}
\vspace*{-1.2truecm}
\caption{Currently allowed region in the $\sigma_d$--$\kappa_d$ plane 
for the (HP+JL)QCD results and $\phi^{\rm NP}_d|_{\rm incl}$.}\label{fig:Bd-HPJLQCD}
\end{figure}

Further constraints are implied by the experimental value of the 
mixing phase $\phi_d= \phi_d^{\rm SM} + \phi_d^{\rm NP}$, which allows us
also to determine contours in the $\sigma_d$--$\kappa_d$ plane, as shown
in Ref.~\cite{BF-DMs}. Interestingly, it turns out that $\kappa_d$ is bounded
from below for any value of $\phi_d^{\rm NP}\not=0$. For example, even a
small phase $|\phi_d^{\rm NP}|=10^\circ$ implies a clean lower bound of
$\kappa_d\geq0.17$, i.e.\ NP contributions of at most 17\%. For the implementation
of these constraints, the NP phase $\phi_d^{\rm NP}$  has to be determined.
To this end, we use the data for the mixing-induced CP violation in 
$B^0_d\to J/\psi K_{\rm S}$ and similar modes, which imply 
$\phi_d=2\beta+\phi_d^{\rm NP}=(43.4\pm2.5)^\circ$. Comparing, on the other hand,
with the ``true" value of $2\beta$ following from $\gamma$ and $R_b$, we obtain
\begin{eqnarray}
\phi^{\rm NP}_d|_{\rm incl} & = & -(10.1\pm 4.6)^\circ\\
\phi^{\rm NP}_d|_{\rm excl} & = & -(2.5\pm 8.0)^\circ
\end{eqnarray}
for the values of $|V_{ub}|$ given in (\ref{Vub-incl}) and (\ref{Vub-excl}). These
values are very stable with respect to $\gamma$, i.e.\ the CKM parameter
dependence is complementary to (\ref{rhod-JLQCD}) and (\ref{rhod-HPJL}). In 
Figs.~\ref{fig:Bd-JLQCD} and \ref{fig:Bd-HPJLQCD}, we show the allowed
regions in NP parameter space arising form $\Delta M_d$ (hills and closed
curves) and $\phi_d$. We see that the impact of the information 
on CP violation is dramatic. On the other hand, values of $\kappa_d$
as large as 0.5 are still allowed.

\boldmath
\subsection{Space for NP in the $B_s$ System}
\begin{figure}
\begin{center}
\includegraphics[width=4.8truecm]{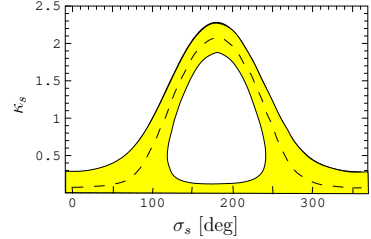}
\end{center}
\vspace*{-1.2truecm}
\caption{Currently allowed region in the $\sigma_s$--$\kappa_s$ plane
following from the $\Delta M_s$ measurement for the JLQCD 
results.}\label{fig:Bs-JLQCD}
\end{figure}

\begin{figure}
\vspace*{-0.4truecm}
\begin{center}
\includegraphics[width=4.8truecm]{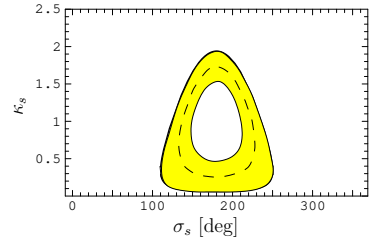}
\end{center}
\vspace*{-1.2truecm}
\caption{Currently allowed region in the $\sigma_s$--$\kappa_s$ plane
following from the $\Delta M_s$ measurement for the (HP+JL)QCD
results.}\label{fig:Bs-HPJLQCD}
\end{figure}

\unboldmath
The analysis of the $B_s$-meson system can be done in analogy to its $B_d$
counterpart, with straightforward replacements of parameters. The relevant
CKM factor can be written, with the help of the unitarity of the CKM matrix, as
$|V_{ts}^*V_{tb}|=|V_{cb}|\left[1+{\cal O}(\lambda^2)\right]$. We then obtain
\begin{eqnarray}
\left.\rho_s\right|_{\rm JLQCD} &=&
1.08^{+0.03}_{-0.01}  \pm 0.19 \\
\left.\rho_s\right|_{\rm  (HP+JL)QCD} &=& 
0.74^{+0.02}_{-0.01} \pm 0.18,
\end{eqnarray}
where the first and second errors have experimental and theoretical origins,
respectively. Consequently, the SM prediction using JLQCD is in perfect 
agreement with the measured value, while the (HP+JL)QCD result is
$1.5\sigma$ larger. A similar pattern, though at the $1\sigma$ level, is interestingly 
also present in (\ref{rhod-JLQCD}) and (\ref{rhod-HPJL}). 
The resulting situation in the $\sigma_s$--$\kappa_s$ plane is shown in 
Figs.~\ref{fig:Bs-JLQCD} and \ref{fig:Bs-HPJLQCD}. We observe that the 
$\Delta M_s$ measurement still leaves a lot of space for the NP parameters.

In the literature, the ratio $\Delta M_s/\Delta M_d$ is often considered. Here
the $SU(3)$-breaking parameter $\xi$ enters, which has a reduced theoretical
uncertainty as compared to $f_{B_q}\hat B_{B_q}^{1/2}$. Usually, the ratio
$|V_{td}/V_{ts}|$ is determined along these lines. If we apply the
unitarity of the CKM matrix, we may, alternatively, determine $\rho_s/\rho_d$
through $\Delta M_s/\Delta M_d$. Because of the currently large range for
$\gamma$, $\rho_s/\rho_d$ is not  stringently constrained, 
although the central values are nicely consistent with 1 for the JLQCD and
(HP+JL)QCD parameters. Even in our
2010 scenario, yielding
\begin{equation}
\left.\frac{\rho_s}{\rho_d}\right|_{2010}=1.07 \pm 0.09
(\gamma,R_b)^{+0.06}_{-0.08}(\xi),
\end{equation}
a stringent test of whether $\rho_s/\rho_d=1$ would still not be possible.

\begin{figure}
\vspace*{-0.4truecm}
\begin{center}
\includegraphics[width=4.8truecm]{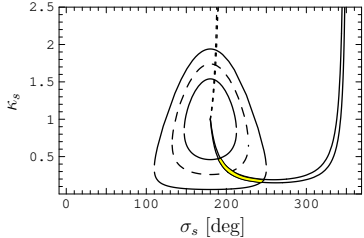}
\end{center}
\vspace*{-1.2truecm}
\caption{Impact of a CP violation measurement on 
the $\sigma_s$--$\kappa_s$ plane in our 2010 scenario.}\label{fig:Bs-2010}
\end{figure}

The golden decay to search for NP effects in $B^0_s$--$\bar B^0_s$ mixing
is $B^0_s\to J/\psi \phi$, the $B_s$ counterpart
of $B^0_d\to J/\psi K_{\rm S}$, allowing us to extract 
\begin{equation}
\sin\phi_s=\sin(-2\lambda^2R_b\sin\gamma+\phi_s^{\rm NP})
\stackrel{\rm SM}{\approx}-0.03
\end{equation}
through mixing-induced CP violation in the time-dependent angular distribution
of its decay products. This decay is very accessible at the LHC. As was noted
in Ref.~\cite{BF-DMs}, even a small NP phase $\phi_s^{\rm NP}\sim -10^\circ$
would yield a CP asymmetry at the $-20\%$ level, so that the possible signal
of NP in $B_d$ mixing could be greatly magnified. Assuming a measurement
of $\sin\phi_s=-0.20\pm0.02$ at the LHC for our 2010 scenario,
we arrive at the situation shown in Fig.~\ref{fig:Bs-2010}. Here the dotted line
refers to $\cos\phi_s<0$, which can also be excluded through further measurements. 
We see that it will be very challenging to establish NP without new CP-violating effects.
On the other hand, Fig.~\ref{fig:Bs-2010} still corresponds to 
$0.2\lsim\kappa_s\lsim 0.5$;  a determination of $\kappa_s$ with $10\%$ accuracy 
would require the reduction of the error of $f_{B_s}\hat B_{B_s}^{1/2}$ to $10\%$, 
i.e. an improvement of the current (HP+JL)QCD lattice results by a factor of 2.

\boldmath
\subsection{Impact of $\Delta M_s$ on NP Scenarios}
\unboldmath
Since $\phi_s$ is still essentially unconstrained, large CP-violating NP effects in 
$B^0_s$--$\bar B^0_s$ mixing may show up at the LHC. Such
effects arise in fact in various extensions of the SM. Let us consider
a model with an extra $Z'$ boson as an example, assuming that $Z$--$Z'$ mixing 
is negligible and that the $Z'$ has flavour non-diagonal couplings only to 
left-handed quarks, which means that its effect is described by only one complex 
parameter: 
\begin{equation}
\rho_L e^{i\phi_L}\equiv\frac{g' M_Z}{g M_{Z'}}\,B_{sb}^L\sim 10^{-3}.
\end{equation}
Translating the $\kappa_s$--$\sigma_s$ constraints into the $\rho_L$--$\phi_L$
space, $\kappa_s<2.5$ implies 
$\rho_L<2.6\times 10^{-3}$, yielding $1.5\,{\rm TeV}(g'/g)|B_{sb}^L/V_{ts}| < M_{Z'}$,
while $\phi_L\leftrightarrow \sigma_s$ leaves $\phi_L$ still essentially unconstrained. 
Also in SUSY scenarios, a lot of space for CP-violating NP effects in 
$B^0_s$--$\bar B^0_s$ mixing is left, as discussed further by V. Lubicz and 
L. Silvestrini.

\section{CONCLUSIONS AND OUTLOOK}
The $B$-decay data agree globally with the Kobayashi--Maskawa picture, but we 
have also hints
for discrepancies. These have to be studied further, and could be first signals of 
NP requiring {\it new} sources of CP violation. The recent measurement
of $\Delta M_s$ at the Tevatron triggered quite some excitement. It still leaves
a lot of space for NP, where a smoking-gun signal would be given  by new 
CP violation in $B^0_s\to J/\psi \phi$ and similar modes, which are very accessible
at the LHC. 

At this new collider, also exciting perspectives for $B$-physics studies 
will emerge, in particular through the full exploitation of the $B_s$ physics potential, various determinations of $\gamma$, and explorations of rare decays such 
as $B_{s,d}\to\mu^+\mu^-$.

Further precision $B$-decay measurements could be performed in the next decade
at an $e^+e^-$ super-$B$ factory. Since also a nice synergy between flavour 
physics and the direct NP searches at ATLAS and CMS is expected, I have no 
doubts that exciting years are ahead of us!

\end{document}